\documentclass[reprint,aps,pra,showpacs,superscriptaddress]{revtex4-1} 
\usepackage{amsmath,amssymb}
\usepackage{mathrsfs}
\usepackage{bbm}
\usepackage[dvipdfmx]{graphicx, color}
\usepackage{dcolumn}
\usepackage{bm}
\usepackage{enumerate}
\usepackage{theorem}
\newtheorem{theorem}{Theorem}
\newtheorem{lemma}{Lemma}
{}

\def\argmin{\mathop{\mathrm{argmin}}}

\def\Trace{\mathop{\mathrm{Tr}}}

\newcommand{\Probability}{\mathbbm{P}}  
\newcommand{\Expectation}{\mathbbm{E}}

\begin{document}
\title{Precision-guaranteed quantum tomography}



\author{Takanori Sugiyama}
\email{sugiyama@eve.phys.s.u-tokyo.ac.jp}
\altaffiliation{\\ Current email address: sugiyama@itp.phys.ethz.ch}
\altaffiliation{\\ Current address: Institute for Theoretical Physics, ETH Zurich
} 
\affiliation{
Department of Physics, Graduate School of Science, The University of Tokyo,\\
7-3-1 Hongo, Bunkyo-ku, Tokyo, Japan 113-0033.
}
\author{Peter S. Turner}
\affiliation{
Department of Physics, Graduate School of Science, The University of Tokyo,\\
7-3-1 Hongo, Bunkyo-ku, Tokyo, Japan 113-0033.
}
\author{Mio Murao}
\affiliation{
Department of Physics, Graduate School of Science, The University of Tokyo,\\
7-3-1 Hongo, Bunkyo-ku, Tokyo, Japan 113-0033.
}
\affiliation{
Institute for Nano Quantum Information Electronics, The University of Tokyo,\\
4-6-1 Komaba, Meguro-ku, Tokyo, Japan 153-8505.
}
\date{\today}
\begin{abstract}
   Quantum state tomography is the standard tool in current experiments for verifying that a state prepared in the lab is close to an ideal target state, but up to now there were no rigorous methods for evaluating the precision of the state preparation in tomographic experiments.
   We propose a new estimator for quantum state tomography, and prove that the (always physical) estimates will be close to the true prepared state with high probability.
   We derive an explicit formula for evaluating how high the probability is for an arbitrary finite-dimensional system and explicitly give the one- and two-qubit cases as examples.
   This formula applies for any informationally complete sets of measurements, arbitrary finite number of data sets, and general loss functions including the infidelity, the Hilbert-Schmidt, and the trace distances. 
   Using the formula, we can evaluate not only the difference between the estimated and prepared states, but also the difference between the prepared and target states. 
   This is the first result directly applicable to the problem of evaluating the precision of estimation and preparation in quantum tomographic experiments.     
\end{abstract}
\pacs{03.65.Wj, 03.67.-a, 02.50.Tt, 06.20.Dk}
\maketitle

 \section{Introduction}
   The development of Science has always been supported by the development of precise and accurate techniques of measurement and control.
   Measurement outcomes are affected by statistical and systematic noise, and evaluating the measurement precision under these errors is a fundamental aspect of those techniques. 
   In Physics, now more than ever high-precision experiments are required for testing whether a theoretical model is suitable for describing Nature. 
   A popular figure of merit for this precision is the standard deviation, but in more demanding experiments a different figure of merit, called a confidence level, is also used. 
   For example, in the search for the standard model Higgs boson, the ATLAS group at the LHC reported an experimental result that narrowed the range of the possible Higgs boson mass at the 95\% confidence level \cite{2012PRD86_ATLAS}.
   This shows the confidence level to be a compelling benchmark for justifying whether an experimental result is reliable or not.

      \begin{figure}[t!]
         \includegraphics[width=0.55\linewidth]{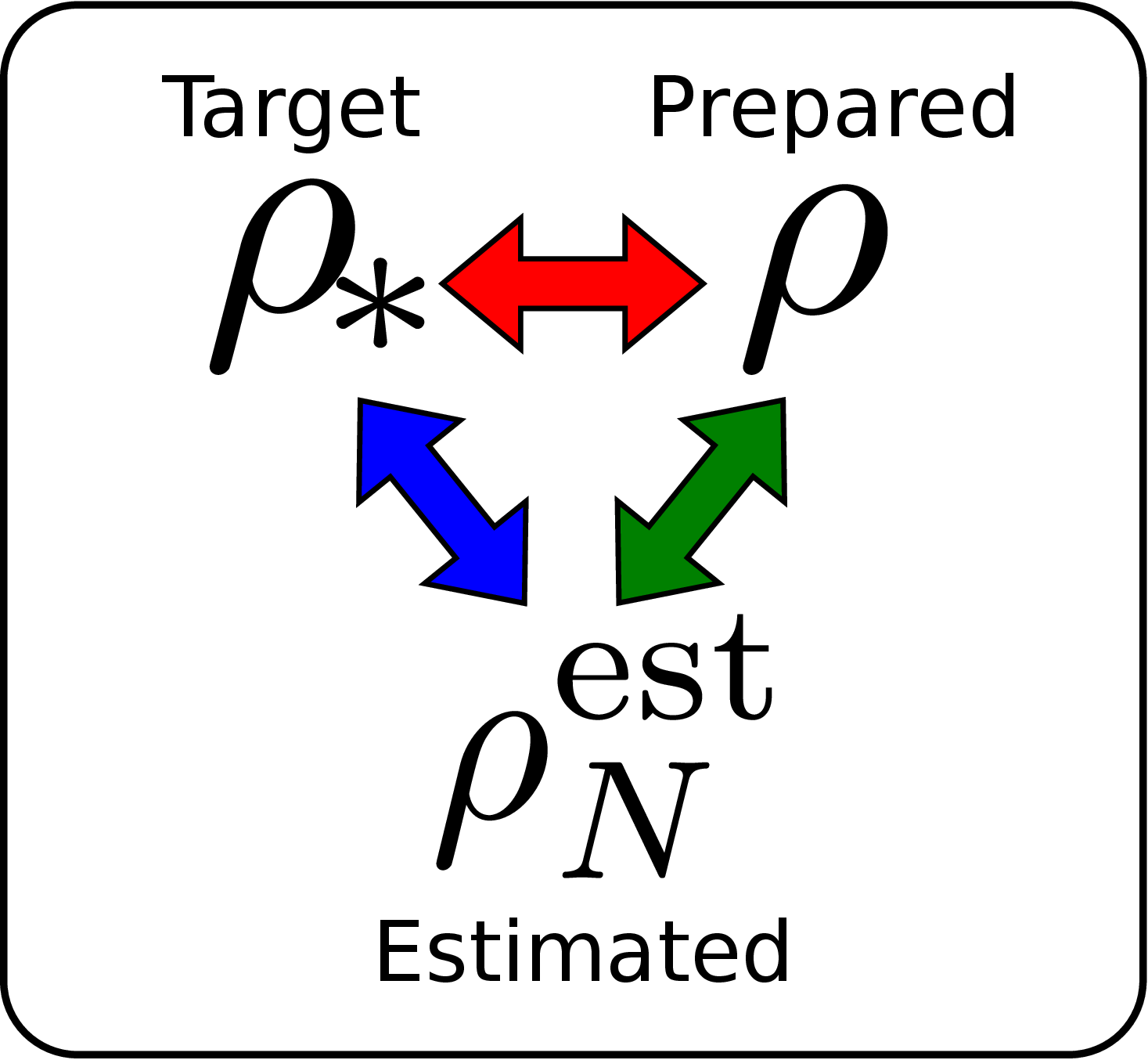}%
         \caption{Three-fold relation between target, prepared, and estimated states: $\rho_{*}$ is a target state that an experimentalist tries to prepare, $\rho$ is the true prepared state. $\rho^{\mathrm{est}}_{N}$ is an estimate made from $N$ tomographic data sets. 
         \label{Fig:1}}
      \end{figure}    

   Quantum information is another field where highly precise measurement and control are necessary.
   It has been shown theoretically that by using ``quantumness", we can perform more efficient computation \cite{Shor1994} and more secure cryptography \cite{BB84} compared to existing protocols.
   In the practical implementation of these new protocols, highly precise preparation and control of specific quantum states are required.
   Quantum tomography is a standard tool in current quantum information experiments for verifying a successful realization of states and operations \cite{QTText2004}.
   Let us consider the case of state preparation, where $\rho_{*}$ denotes a target state that we are trying to prepare in the lab.
   In real experiments, the true prepared state $\rho$ does not coincide with $\rho_{*}$ because of imperfections.
   We wish to evaluate the precision of this preparation, that is, the difference between $\rho_{*}$ and $\rho$ - however we do not know $\rho$.
   Instead, we perform quantum state tomography;  let $\rho^{\mathrm{est}}_{N}$ denote an estimate of the state made from $N$ sets of data obtained in a tomographic experiment.
   To date the best we have been able to do is to evaluate the difference between $\rho_{*}$ and $\rho^{\mathrm{est}}_{N}$ (see Fig. \ref{Fig:1}), but even if the difference is small,  it does not guarantee that the prepared state $\rho$ is close to the target state $\rho_{*}$, because $\rho^{\mathrm{est}}_{N}$ is given probabilistically and can deviate from $\rho$ when $N$ is finite. 
   In this context, we refer to the difference between $\rho^{\mathrm{est}}_{N}$ and $\rho$ the {\it precision} of the estimation.

   There are many proposals for evaluating the precision of estimation \cite{Robin2012,ChristandlRenner2012PRL,Flammia2012NJP} and preparation \cite{Stieve2011_PRL106,Silva2011PRL107}.   
   An approach using confidence regions is one currently popular example.
   Unlike standard quantum tomography, in this approach the estimate is not a point but a region in state space.
      In \cite{Robin2012,ChristandlRenner2012PRL}, confidence region estimators for quantum state estimation were proposed, and their volume of the region were analyzed.    
      The confidence level can be used for evaluating the precision of region estimates, but these cannot be directly applied for evaluating the precision of point estimates in tomographic experiments and state preparation.
      In \cite{Flammia2012NJP}, two state estimators were proposed that use random sampling of Pauli measurements.
      Called compressed sensing, the authors proved that the estimates are close to the true state with high probability.
      However, the formulae derived for evaluating the difference between the estimates and the true state include indeterminate coefficients, and the value of the difference cannot be calculated.     
      Therefore the compressed sensing results are not directly applicable to the evaluation of the estimation precision of tomographic experiments.
      In \cite{Stieve2011_PRL106,Silva2011PRL107}, a method for estimating the difference between $\rho_{*}$ and $\rho$ as evaluated by the fidelity was proposed.
      Called direct fidelity estimation, it avoids point estimates altogether, and by performing random Pauli measurements it allows the precision of state preparation to be evaluated, (it assumes that the target state $\rho_{*}$ is pure).
      However, the method is not capable of evaluating the estimation precision of point estimates in quantum tomography.
      Thus, this remains a crucial problem in the current theory of quantum tomography.

    Here we give a solution to this problem.
    We propose a new point estimator for quantum state tomography in finite-dimensional systems, and prove that the estimated states are within a distance threshold $\delta$ from the prepared state with high probability.
    We derive an explicit formula for evaluating how high the probability is in arbitrary finite-dimensional systems.     
    This formula applies for any informationally complete set of measurements, for an arbitrary finite amount of data, and for general loss functions including the infidelity, the Hilbert-Schmidt, and the trace distances. 
   Importantly, for a given experimental setup we can calculate the value of the formula without knowing the true prepared state, and so the formula can be used to evaluate the precision of state preparation.
   To our knowledge this is the first result directly applicable to evaluating the precisions of both estimation and preparation in quantum tomography.
   We demonstrate the technique for the example of one- and two-qubit state tomography.

\section{Preliminaries}
      We consider a finite $d (<\infty )$ dimensional quantum system, with Hilbert space $\mathcal{H}$.
      A state of the system is described by a density matrix, which is a positive-semidefinite and trace-one matrix, the space of which we denote by $\mathcal{S}(\mathcal{H})$.   
      Let $\rho$ denote the density matrix describing the true prepared state on $\mathcal{H}$.
      It is unknown, and we make no further assumptions on $\rho$.
      Suppose that $N$ identical copies of the unknown true state, $\rho^{\otimes N}$, are available, and we can perform a measurement on each copy.      
      Our aim is to estimate the unknown true state $\rho$ from measurement results. 
      Let $\bm{\lambda} = (\lambda_{1}, \ldots , \lambda_{d^{2} - 1})$ denote a set of Hermitian matrices satisfying (i) (Tracelessness) $\Trace [\lambda_{\alpha}] = 0$ and (ii) (Orthogonality) $\Trace [\lambda_{\alpha} \lambda_{\beta}] = 2\delta_{\alpha \beta}$. 
       Using this set, a density matrix can be parametrized as \cite{Kimura03,Byrd03}
       \begin{eqnarray}
          \rho (\bm{s}) &=& \frac{1}{d} I + \frac{1}{2} \bm{s} \cdot \bm{\lambda}, \label{eq:gBloch_state}
       \end{eqnarray}
       where $I$ is the identity matrix and $\bm{s}$ is a vector in $\mathbbm{R}^{d^{2} - 1}$.
       The matrix and vector are uniquely related as $s_{\alpha} = \Trace [\rho \lambda_{\alpha}]$.
       Positive-semidefiniteness of $\rho$ requires constraints on the parameter space.
       Let $S$ denote the set of parameters corresponding to density matrices, and $S$ is a convex subset of $\mathbbm{R}^{d^{2}-1}$.    
       Estimation of $\rho \in \mathcal{S}(\mathcal{H})$ is equivalent to that of $\bm{s} \in S$.    
       
       The statistics of a quantum measurement are described by a positive operator-valued measure (POVM), which is a set of positive-semidefinite matrices that sum to the identity.
       In standard setting of quantum tomography we choose a combination of measurements.
       Let $\vec{\bm{\Pi}} = \{ \bm{\Pi}^{(j)} \}_{j=1}^{J}$ denote a finite set of POVMs.               
       Suppose that for estimating $\rho$ we independently perform a measurement described by a POVM $\bm{\Pi}^{(j )}= \{ \Pi^{(j )}_{m} \}_{m=1}^{M^{(j)}}$ a number $n^{(j )}$ of times ($j= 1, \ldots , J$).
       The total number of measurement trials is $\sum_{j=1}^{J}n^{(j)} = N$.
       Let us define $r^{(j)}:= N / n^{(j)}$.               
       Elements of the POVM can also be parametrized as
       \begin{eqnarray}   
         \Pi^{(j )}_{m} &=& a^{(j )}_{m, 0} I + \bm{a}^{(j )}_{m} \cdot \bm{\lambda}, \label{eq:gBloch_POVM}           
       \end{eqnarray}
       where $a^{(j )}_{m, 0} = \Trace [ \Pi^{(j )}_{m} ] / d$, and $a^{(j )}_{m,\beta}= \Trace [ \Pi^{(j )}_{m} \lambda_{\beta} ] /2$.        
       When we perform the measurement on a system in a state described by $\rho$, the probability that we observe an outcome $m$ is given by
       \begin{eqnarray}
          p(m | \bm{\Pi}^{(j)}, \rho)         
             = \Trace \left[ \Pi^{(j )}_{m} \rho \right] 
             = a^{(j )}_{m, 0} + \bm{a}^{(j )}_{m} \cdot \bm{s}.\label{eq:BornRule_gBloch_1}
        \end{eqnarray}    
        A set of POVMs $\vec{\bm{\Pi}} := \left\{ \bm{\Pi}^{(j)} \right\}_{j = 1}^{J}$ is called informationally complete (IC) if it spans the vector space of Hermitian matrices on $\mathcal{H}$ \cite{Prugovecki77}.    
        Such a set allows for the reconstruction of an arbitrary quantum state, and we will assume that our $\vec{\bm{\Pi}}$ is always IC.

      Let $n^{(j)}_{m}$ denote the number of appearances of outcome $m$ in the data from the $n^{(j)}$ measurement trials of $\bm{\Pi}^{(j)}$ ($m=1, \ldots, M^{(j)}$); then $f^{(j)}_{m}:= n^{(j)}_{m} / n^{(j)}$ is the relative frequency.       
      A map from a data set to the space of interest -- in this case the space of quantum states -- is called an estimator, and an estimation result is called an estimate.
     One of the simplest is a linear estimator, $\rho^{\mathrm{L}}$ \cite{Fano1957}, defined as a matrix $\sigma$ satisfying 
     \begin{eqnarray}
         f^{(j)}_{m} = p(m| \bm{\Pi}^{(j)}, \sigma ),\ j = 1, \ldots , J,\ m=1, \ldots , M^{(j)}. \label{def:LinearEstimator}
      \end{eqnarray}
      The idea is to use the relative frequencies instead of the unknown true probability distributions.
      This might seem natural to physicists, but there are two problems.
      The first problem is that when $\vec{\bm{\Pi}}$ is over-complete, equation (\ref{def:LinearEstimator}) might have no solutions, {\it i.e.}, we happen to obtain a data set from which we cannot calculate the estimate.
      The second problem is that even if there exists a solution of equation (\ref{def:LinearEstimator}), the solution can be unphysical, {\it i.e.}, lie outside of $S$. 
      For these two reasons linear estimators are rarely used in tomographic experiments anymore.     
      The current standard is a maximum-likelihood (ML) estimator, which is defined as the point in $S$ maximizing the likelihood function \cite{Hradil1997}.
      By definition, such estimates are always physical.
      The asymptotic ($N \sim \infty$) behavior of the confidence level of a ML estimator is analyzed in \cite{Sugiyama2011}, but there have been no such results for finite data sets.

      A loss function is a measure for evaluating the difference between two states.
      We analyze the following three loss functions:    
      \begin{eqnarray}
         \Delta^{\mathrm{HS}} (\rho^{\prime}, \rho) &:=& \frac{1}{\sqrt{2}} \Trace \left[ ( \rho^{\prime} - \rho )^{2} \right]^{1/2}, \\
         \Delta^{\mathrm{T}} (\rho^{\prime}, \rho) &:=& \frac{1}{2} \Trace \left[ | \rho^{\prime} - \rho | \right], \\
         \Delta^{\mathrm{IF}} (\rho^{\prime}, \rho) &:=& 1 - \Trace \left[ \sqrt{ \sqrt{\rho^{\prime}} \rho \sqrt{\rho^{\prime}}  } \right]^{2}.   
      \end{eqnarray}
      called the Hilbert-Schmidt distance, the trace distance, and the infidelity, respectively.
      In current quantum tomography experiments the trace distance and the infidelity are most often used.

\section{Results}    
      Instead of ML estimator, we propose a new estimator $\rho^{\mathrm{ENM}}$.
      In order to define the new estimator, we introduce a different estimator $\rho^{\mathrm{LLS}}$.
      Let us define $\bm{p} ( \sigma)$ and $\bm{f}_{N}$ as vectors with $(j,m)$-th element $p(m|\bm{\Pi}^{(j)}, \sigma)$ and $f^{(j)}_{m}$, respectively.
      The estimate of $\rho^{\mathrm{LLS}}$ is defined as 
      \begin{eqnarray}
         \rho^{\mathrm{LLS}}_{N} 
            := \argmin_{\sigma; \sigma = \sigma^{\dagger}, \atop \ \ \ \Trace [\sigma ] =1} 
             \| \bm{p} (\sigma ) - \bm{f}_{N}\|_{2}. \label{eq:LLS_matrix}
      \end{eqnarray}
      The range of the minimization in equation (\ref{eq:LLS_matrix}) is restricted by the Hermiticity and trace-one condition, but the positive-semidefiniteness ($\sigma \ge 0$) is not required.
      Therefore the estimates can be unphysical.       
      $\rho^{\mathrm{LLS}}_{N}$ is a linear least squares (LLS) estimator in statistics \cite{LinearModelsAndGeneralizations2008}.
      Let us define $\bm{a}_{0}$ as a vector with $(j,m)$-th element $a^{(j)}_{m, 0}$ and $A$ as a matrix with $[(j,m), \alpha ]$-th element $a^{(j)}_{m, \alpha}$ ($\alpha = 1, \ldots , d^{2} -1$).
      When $\vec{\bm{\Pi}}$ is IC, $A$ is full-rank, and the left-inverse matrix exists and is given by $A^{-1}_{\mathrm{L}} = ( A^{T} A )^{-1} A^{T}$.       
      Then the minimization in equation (\ref{eq:LLS_matrix}) has the analytical solution, and the LLS estimate of the Bloch vector, $\bm{s}^{\mathrm{LLS}}_{N}$, is given as
      \begin{eqnarray}
         \bm{s}^{\mathrm{LLS}}_{N} = A^{-1}_{\mathrm{L}} (\bm{f}_{N} - \bm{a}_{0} ). \label{eq:sLLS_ExplicitForm}
      \end{eqnarray}
      The LLS estimate of density matrix is calculated by $\rho^{\mathrm{LLS}}_{N} = \rho (\bm{s}^{\mathrm{LLS}}_{N})$.
      Using the LLS estimator, we define the estimate of a new estimator, $\rho^{\mathrm{ENM}}$, as
      \begin{eqnarray}
         \rho^{\mathrm{ENM}}_{N} 
            := \argmin_{\rho^{\prime} \in \mathcal{S}(\mathcal{H})} \| \rho^{\prime} - \rho^{\mathrm{LLS}}_{N} \|_{2}.\label{def:ENM_state}
      \end{eqnarray}
      We call $\rho^{\mathrm{ENM}}$ an extended norm-minimization (ENM) estimator \footnote{The ENM estimator is categorized into a weighted least squares estimator with nonlinear constraints.}.
      The estimates are again always physical by definition.

      The following theorem establishes that the ENM estimates are close to the true prepared state with high probability.
      \begin{theorem}[Confidence level of ENM estimator]\label{theorem:ConfLevel_CLS_state}         
         Suppose that $\vec{\bm{\Pi}}$ is IC.
         Then for arbitrary true density matrix $\rho \in \mathcal{S} (\mathcal{H})$, set of positive integers $n^{(j)}$ satisfying $\sum_{j=1}^{J}n^{(j)} = N$, and positive number $\delta$,
         \begin{eqnarray}
            \Delta (\rho^{\mathrm{ENM}}_{N} , \rho) \le \delta \label{eq:ErrorThrethold_CLS}
         \end{eqnarray}
         holds with probability at least
         \begin{eqnarray}
            CL:=1 - 2 \sum_{\alpha = 1}^{d^{2} - 1} \exp \left[- \frac{b}{c_{\alpha}}\delta^{2} N \right],\label{eq:ConfidenceLevel_CLS}
         \end{eqnarray}
         where $b$ is determined by our choice of the loss function as 
         \begin{eqnarray}
            b &:=& \left\{
                   \begin{array}{ll}
                     8 / (d^{2} - 1)    & \mbox{if}\ \Delta = \Delta^{\mathrm{HS}} \\
                     16 / d (d^{2} - 1) & \mbox{if}\ \Delta = \Delta^{\mathrm{T}}\\ 
                     4 / d (d^{2} - 1) & \mbox{if}\ \Delta = \Delta^{\mathrm{IF}}                
                   \end{array}
                   \right. , \label{eq:MainTheorem_CLS_b} 
         \end{eqnarray}
         and $c_{\alpha}$ are determined by our choice of the measurement setting as 
         \begin{eqnarray}
            c_{\alpha} := \sum_{j=1}^{J} r^{(j)} 
                               \left\{ \max_{m} \left[A^{-1}_{\mathrm{L}} \right]_{\alpha , (j, m)} 
                                - \min_{m} \left[ A^{-1}_{\mathrm{L}} \right]_{\alpha , (j, m)} \right\}^{2}.
         \end{eqnarray}    
      \end{theorem}    
      We call $CL$ in equation (\ref{eq:ConfidenceLevel_CLS}) the confidence level of $\rho^{\mathrm{ENM}}$ at the (user-specified) error threshold $\delta$.

      We sketch the proof here, with the details shown in Appendix.
      The proof consists of two steps.
      First, we consider the Hilbert-Schmidt distance case and analyze the probability that we obtain estimates deviating from the true density matrix by more than the error threshold $\delta$.
      We call this probability the estimation error probability with respect to $\Delta = \Delta^{\mathrm{HS}}$ at the error threshold $\delta$.
      By using known results in convex analysis, we prove that the estimation error probability of $\rho^{\mathrm{ENM}}$ is smaller than that of $\rho^{\mathrm{LLS}}$.
      Second, we  derive an upper-bound on the estimation error probability of $\rho^{\mathrm{LLS}}$.
      In the derivation, we reduce the analysis of the probability for multi-parameter estimation to that of the probability for one-parameter estimation and derive a $\rho$-independent upper-bound of the probability with the Hoeffding's tail inequality \cite{Hoeffding1963}.
      From the result of the first step, the derived upper-bound is also an upper-bound of that of $\rho^{\mathrm{ENM}}$.  
      The confidence level of $\Delta^{\mathrm{HS}}$ is given by one minus the upper-bound of the estimation error probability.      
      The confidence levels of $\Delta^{\mathrm{T}}$ and $\Delta^{\mathrm{IF}}$ are derived by combining that of $\Delta^{\mathrm{HS}}$ with inequalities between these loss functions.

      \begin{figure*}[t]
         \includegraphics[width=0.90\linewidth]{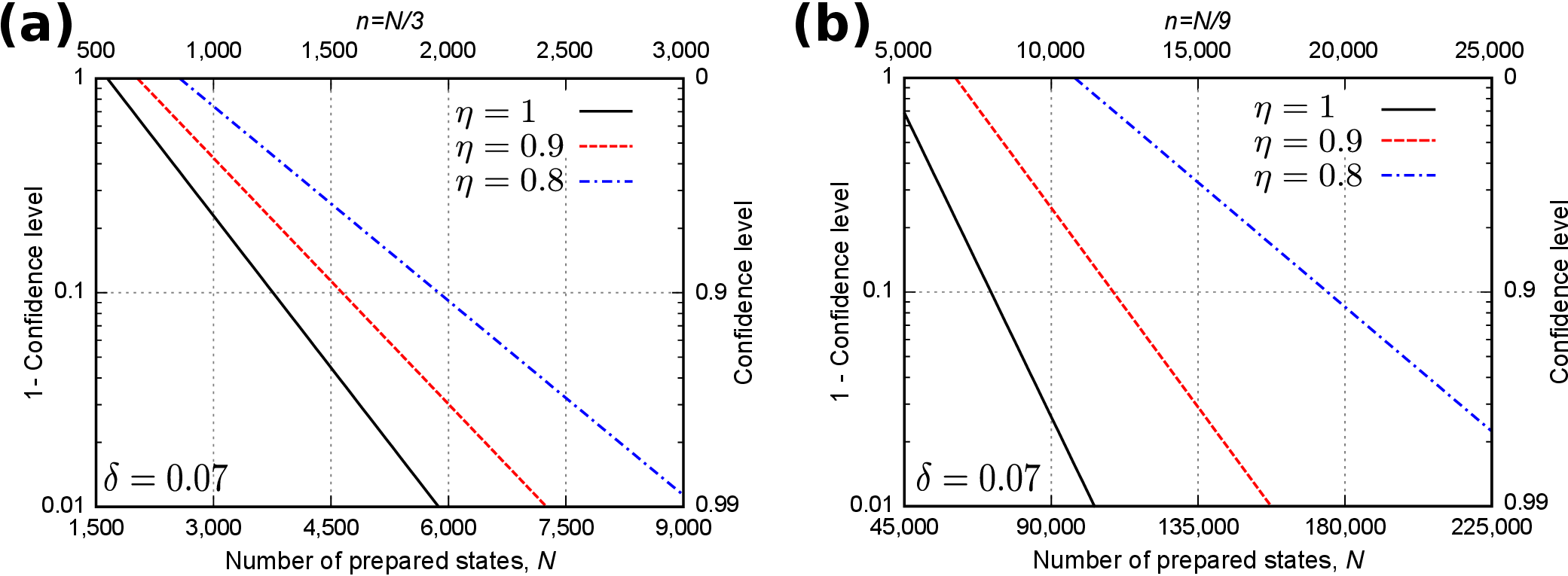}%
         \caption{Confidence level of $\rho^{\mathrm{ENM}}$ for error threshold $\delta = 0.07$ in quantum state tomography: panel (a) is the one-qubit case ($k=1$) and panel (b) is the two-qubit case ($k=2$). In both panels, the left and right vertical axes are $1 - CL (k)$ and $CL(k)$ in equation (\ref{eq:CL_kqubit_state}), respectively. The lower and upper horizontal axes are the number of prepared states $N$ and the number of observations for each tensor product of Pauli matrices $n= N / 3^{k}$, respectively. In both panels, the line styles are fixed as follows: solid (black) line for detection efficiency $\eta = 1$, dashed (red) line for $\eta = 0.9$, chain (blue) line for $\eta = 0.8$.  \label{Fig2:OneTwo}}
      \end{figure*}

\section{Analysis}
      The most important point in Theorem \ref{theorem:ConfLevel_CLS_state} is that equation (\ref{eq:ConfidenceLevel_CLS}) is independent of the true prepared state $\rho$.
      Therefore we can use it to evaluate $\Delta (\rho_{*}, \rho)$ without knowing $\rho$.
      Suppose that we choose a loss function $\Delta$ that satisfies properties of a mathematical distance.
      Then from the triangle inequality and Theorem \ref{theorem:ConfLevel_CLS_state}, we have
      \begin{eqnarray}
         \Delta (\rho_{*}, \rho) 
           &\le& \Delta (\rho_{*}, \rho^{\mathrm{ENM}}_{N}) + \Delta (\rho^{\mathrm{ENM}}_{N}, \rho) \label{eq:TriangleIneq_1}\\
           &\le& \Delta (\rho_{*}, \rho^{\mathrm{ENM}}_{N}) + \delta , \label{eq:Triangle2}
      \end{eqnarray}       
      where equation (\ref{eq:Triangle2}) holds at the confidence level in equation (\ref{eq:ConfidenceLevel_CLS}).
      We can calculate the value of the R.H.S. of equation (\ref{eq:Triangle2}) without knowing the true prepared state $\rho$ and use it to evaluate the size of $\Delta (\rho_{*}, \rho )$.   
       In tomographic experiments, the infidelity, (or the fidelity $F(\rho^{\prime}, \rho ):= 1 - \Delta^{\mathrm{IF}} (\rho^{\prime}, \rho )$), is a popular loss function used for evaluating $\Delta (\rho_{*}, \rho^{\mathrm{est}}_{N})$.
       The infidelity is not a distance, it does not satisfy the triangle inequality, but it is related to the trace distance by $\Delta^{\mathrm{IF}} (\rho^{\prime}, \rho ) \le 2 \Delta^{\mathrm{T}} (\rho^{\prime}, \rho)$ \cite{GeometryofQuantumStates2006}.
       Thus we obtain
       \begin{eqnarray}
          \Delta^{\mathrm{IF}} ( \rho_{*}, \rho) 
          &\le& 2 \{  \Delta^{\mathrm{T}} (\rho_{*}, \rho^{\mathrm{ENM}}_{N}) +  \delta \},  \label{eq:ErrorThrethold_IF_T}  \\   
          F(\rho_{*}, \rho ) &\ge& 1 - 2 \{  \Delta^{\mathrm{T}} (\rho_{*}, \rho^{\mathrm{ENM}}_{N}) +  \delta \}, \label{eq:ErrorThrethold_F}              
       \end{eqnarray}
       where Eqs. (\ref{eq:ErrorThrethold_IF_T}) and (\ref{eq:ErrorThrethold_F}) hold at the confidence level in equation (\ref{eq:ConfidenceLevel_CLS}) for $\Delta^{\mathrm{T}}$.

   Let us consider quantum state tomography of a $k$-qubit system and suppose that we make the three Pauli measurements with detection efficiency $\eta$ on each qubit.
   There are $3^{k}$ different tensor products of Pauli matrices ($J=3^{k}$), and suppose that we observe each equally $n:= N / 3^{k}$ times.
   Let us choose $\bm{\lambda}$ to be the set of tensor products of Pauli and identity matrices with the normalization factor $1 / \sqrt{2^{k-1}}$.
   From the relation $s_{\alpha} = \Trace [ \rho \lambda_{\alpha} ]$, we obtain $c_{\alpha} = 2^{3-k} \cdot 3^{k-l} / \eta^{2(k-l)}$, where $l = 0, \ldots , k-1$ is the number of the identity matrices appearing in $\lambda_{\alpha}$.
   From the information above, we can derive the explicit form of $CL$ for $k$-qubit state tomography.
   When we choose $\Delta = \Delta^{\mathrm{T}}$, we have
   \begin{eqnarray}
      &&CL(k) \notag \\
      &&= 1 - 2 \sum_{l=0}^{k-1} 3^{k-l}  \binom{k}{l} \exp \left[ - \frac{2}{2^{2k} - 1} \frac{\eta^{2(k-l)}}{3^{k-l}} \delta^{2} N \right]. \label{eq:CL_kqubit_state}
   \end{eqnarray}
   The details of the derivation of equation (\ref{eq:CL_kqubit_state}) are explained in Appendix.
   Figure \ref{Fig2:OneTwo} shows plots of equation (\ref{eq:CL_kqubit_state}) for the one-qubit ($k=1$) and two-qubit ($k=2$) cases in panels (a) and (b), respectively. 
   The error threshold is $\delta= 0.07$ and detection efficiency is $\eta = 1,\  0.9,\  0.8$.
   Both panels indicate that smaller detection efficiency requires a larger number of prepared states.
   The plots tell us what value of $N$ sufficient for guaranteeing a fixed confidence level.
   For example, if we want to guarantee $99\%$ confidence level for $\delta=0.07$ in one-qubit state tomography with $\eta = 0.9$, panel (a) indicates that $N=7,500$ is sufficient for that.

      In \cite{SmolinPRL108_2012}, an efficient ML estimator algorithm is proposed for quantum state tomography using an IC set of projective measurements with Gaussian noise whose variance is known, and numerical results for $k$-qubit ($k=1,\ \ldots , 9$) state tomography indicate that the computational cost would be significantly lower than that of standard ML algorithms.
      In general, a ML estimator is different from the ENM estimator, but in the setting considered in \cite{SmolinPRL108_2012} the ML estimator is a specific case of an ENM estimator, which is defined for general IC measurements.  
      Despite this generality, we find that their efficient algorithm can be modified and used for our ENM estimates \footnote{These calculations involve numerical errors, and strictly speaking the calculated estimate is different from the exact ENM estimate. We analyze the effect of numerical (and systematic) errors on error thresholds and confidence levels and give a solution to this problem in Supplemental Information.}.
       Additionally, our result (Theorem \ref{theorem:ConfLevel_CLS_state}) shows that the ENM estimator can be used without assuming projective measurements or Gaussian noise.

     It is natural to ask if instead of performing two sequential optimizations as in the ENM case one performs a single constrained optimization.  
     This is well-known in classical statistics and was applied to a quantum estimation problem in \cite{Tan1997}.  
     Define a constrained least squares estimator
      \begin{eqnarray}
         \rho^{\mathrm{CLS}}_{N} := \argmin_{\rho^{\prime} \in \mathcal{S}(\mathcal{H})} \| \bm{p} (\rho^{\prime}) - \bm{f}_{N} \|_{2}, \  \label{eq:CLS_def}        
      \end{eqnarray} 
      which always exists and is always physical.
      Using nearly the same proof as in Theorem 1, we can derive a confidence level for $\rho^{\mathrm{CLS}}$.
      The result is equivalent or smaller than Eq. (\ref{eq:ConfidenceLevel_CLS})---the details and a comparison to $\rho^{\mathrm{ENM}}$ are shown in Appendix E.
      Although in some cases the confidence levels coincide, in order to calculate CLS estimates we need to solve the quadratic optimization problem under inequality constraints, which the ENM case avoids.

\section{Summary}
      We considered quantum state tomography in arbitrary finite dimensional systems, proposing a new point estimator and deriving an explicit formula evaluating the precision of estimation for an arbitrary finite number of measurement trials.
      We applied the idea using as an example $k$-qubit state tomography with detection errors, and showed plots for the one- and two-qubit cases.
      We also show how the formula can be used for evaluating the precision of state preparation, i.e.,  the difference between the state that we want to prepare and the state that we actually prepared.
      To the best of our knowledge this is the first result that makes it possible to evaluate the precision of estimation and preparation without knowing the prepared state, and we hope it finds application in the analysis of experimental data.

\begin{acknowledgments} 
   The authors thank the Referee for drawing their attention to \cite{Tan1997}.
   TS thanks Steven T. Flammia, David Gross, and Fuyuhiko Tanaka for helpful comments on least squares estimators.
   This work was supported by the JSPS Research Fellowships for Young Scientists (22-7564), the JSPS Postdoctoral Fellowships for Research Abroad (H25-32), and the Project for Developing Innovation Systems of the Ministry of Education, Culture, Sports, Science and Technology (MEXT), Japan.
\end{acknowledgments}


\newpage

\appendix

\begin{center}
{\LARGE {\bf Appendix}}\\
\end{center}

    In Appendix A, we give the proof of Theorem 1, explain the generalization to quantum process tomography, and discuss the possible improvement of Theorem 1.    
    In Appendix B, we derive the confidence level for $k$-qubit state tomography using the Pauli measurements with detection losses.
    In Appendix C and D, we explain a way of evaluating the effect of systematics and numerical errors on an error threshold and confidence level.
    In Appendix E, we derive a confidence level of the CLS estimator and compare the performance to that of the ENM estimator.

\section{Proof of Theorem 1}

      In order to prove Theorem 1 in the manuscript, we introduce four lemmas.
      Lemmas \ref{lemma:Non-expandability} and \ref{lemma:HoeffdingTailIneq} are used in proving Lemmas \ref{lemma:ErrorProb_ENM_LLS} and \ref{lemma:ErrorProb_LLS_LInfty}, respectively.

      First, we introduce a lemma about projections in Euclidean space, which is known in convex analysis and so we omit the proof.
       This lemma is used for proving that the ENM estimator gives more precise estimates than those of the LLS estimator. 
       \begin{lemma}[Non-expandability of projection \cite{BorweinLewis_text2006}] \label{lemma:Non-expandability}       
         Let $K$ denote a positive integer and $V$ denote a non-empty closed convex subset of $\mathbbm{R}^{K}$. 
         In convex analysis, a vector $\bm{v}:=\argmin_{\bm{v}^{\prime}\in V} \| \bm{v}^{\prime} - \bm{w} \|_{2}$ is called the projection of $\bm{w} \in \mathbbm{R}^{K}$ into $V$, and it is known that for any $\bm{v}^{\prime\prime} \in V$, 
         \begin{eqnarray}
            \| \bm{v} - \bm{v}^{\prime\prime} \|_{2} \le \| \bm{w} - \bm{v}^{\prime\prime} \|_{2} \label{eq:NonExpandability}
         \end{eqnarray}
         holds.
     \end{lemma}

            Second, by using Lemma \ref{lemma:Non-expandability}, we prove that the ENM estimator gives more precise estimates than those of the LLS estimator.            
            Let us define $\Probability \left[ \Delta (\rho^{\mathrm{est}}_{N}, \rho) > \delta \right]$ as the probability that we obtain estimates deviating from the true density matrix by more than $\delta$ with respect to a loss function $\Delta$.
            We call $\Probability \left[ \Delta (\rho^{\mathrm{est}}_{N}, \rho) > \delta \right]$ the estimation error probability of estimator $\rho^{\mathrm{est}}$ with respect to loss function $\Delta$.
            Then the following lemma holds:
            \begin{lemma}\label{lemma:ErrorProb_ENM_LLS}
               Suppose that $\vec{\bm{\Pi}}$ is IC.
               For arbitrary $\rho \in \mathcal{S} (\mathcal{H})$, set of positive integers $n^{(j)}$ satisfying $\sum_{j=1}^{J}n^{(j)} = N$, and positive number $\delta$,   
               \begin{eqnarray}
                  \Probability \left[ \Delta^{\mathrm{HS}} (\rho^{\mathrm{ENM}}_{N}, \rho) > \delta \right]
                  \le \Probability \left[ \Delta^{\mathrm{HS}} (\rho^{\mathrm{LLS}}_{N}, \rho) > \delta \right] \label{eq:Lemma2}   
               \end{eqnarray}
               holds.
            \end{lemma}
            \textbf{Proof} (Lemma \ref{lemma:ErrorProb_ENM_LLS})
               The Hilbert-Schmidt distance and $\ell_{2}$-distance on $\mathbbm{R}^{d^{2}-1}$ in the Bloch representation are related by $\Delta^{\mathrm{HS}} (\rho^{\prime}, \rho) = \frac{1}{2} \| \bm{s}^{\prime} - \bm{s}  \|_{2}$.
               The definition of $\rho^{\mathrm{ENM}}$ is rewritten in the vector representation as
               \begin{eqnarray}
                  \bm{s}^{\mathrm{ENM}}_{N}
                     := \argmin_{\bm{s}^{\prime} \in S} \| \bm{s}^{\prime} - \bm{s}^{\mathrm{LLS}}_{N} \|_{2}. \label{def:ENM_vector}
               \end{eqnarray}
               Equation (\ref{def:ENM_vector}) indicates that $\bm{s}^{\mathrm{ENM}}_{N}$ is the projection of $\bm{s}^{\mathrm{LLS}}_{N}$ into the convex subset $S$, and by substituting $K=d^{2} -1$, $\bm{v} = \bm{s}^{\mathrm{ENM}}_{N}$, $\bm{w} = \bm{s}^{\mathrm{LLS}}_{N}$, and $\bm{v}^{\prime\prime} = \bm{s}$ into equation (\ref{eq:NonExpandability}), we obtain $\| \bm{s}^{\mathrm{ENM}}_{N} - \bm{s} \|_{2} \le \| \bm{s}^{\mathrm{LLS}}_{N} - \bm{s} \|_{2}$.
               Then we have
               \begin{eqnarray}
                  \| \bm{s}^{\mathrm{ENM}}_{N} - \bm{s} \|_{2} > 2\delta
                  \Longrightarrow 
                  \| \bm{s}^{\mathrm{LLS}}_{N} - \bm{s} \|_{2} > 2\delta,                
               \end{eqnarray}
               and equation (\ref{eq:Lemma2}) holds.                    
            $\blacksquare$  \\          
            Lemma \ref{lemma:ErrorProb_ENM_LLS} indicates that the estimation error probability of $\rho^{\mathrm{ENM}}$ with respect to $\Delta^{\mathrm{HS}}$ is always smaller than or equivalent to that of $\rho^{\mathrm{LLS}}$.

         Third, we introduce a lemma about an error probabilities of a sum of independent random variables, which is known in statistics and we omit the proof.
         \begin{lemma}[Hoeffding's tail probability inequality\cite{Hoeffding1963}]\label{lemma:HoeffdingTailIneq}
              Let $X_{1}, \ldots , X_{N}$ be independent bounded random variables such that $X_{i}$ takes value in $[ t_{i}, u_{i} ]$ with probability one.
              Let us define $S_{N}:= \sum_{i=1}^{N} X_{i}$.
              Then for any $\delta > 0$ we have
              \begin{eqnarray}
                \hspace{-3mm} \Probability \left[ | S_{N} - \Expectation [S_{N}]  | > \delta \right] \le 2 \exp \left[ -2\delta^{2} / \sum_{i=1}^{N} (u_{i} - t_{i} )^{2} \right], \label{eq:HoeffdingTailIneq}
              \end{eqnarray}
              where $\Expectation [ S_{N} ]$ is the expectation of the random variable $S_{N}$.              
         \end{lemma}
         Lemma \ref{lemma:HoeffdingTailIneq} is used for deriving an upper-bound on an estimation error probability of $\bm{s}^{\mathrm{LLS}}$.

            Fourth, by using Lemma \ref{lemma:HoeffdingTailIneq}, we derive an upper-bound on the estimation error probability of $\bm{s}^{\mathrm{LLS}}$ with respect to the $\ell_{\infty}$-distance.            
            \begin{lemma}\label{lemma:ErrorProb_LLS_LInfty}
               For arbitrary $\bm{s} \in S$, set of positive integers $n^{(j)}$ satisfying $\sum_{j=1}^{J}n^{(j)} = N$, and positive number $\delta$,            
               \begin{eqnarray}
                  \Probability \left[ \| \bm{s}^{\mathrm{LLS}}_{N} - \bm{s} \|_{\infty} > \delta \right] 
                  \le 2 \sum_{\alpha = 1}^{d^{2} - 1} \exp \left[ - \frac{2}{c_{\alpha}} \delta^{2} N \right] \label{eq:ErrorProb_LLS_LInfty}
               \end{eqnarray}   
               holds where
               \begin{eqnarray}
                  \hspace{-4mm}c_{\alpha} := \sum_{j=1}^{J} r^{(j)} 
                               \left\{ \max_{m} \left[A^{-1}_{\mathrm{L}} \right]_{\alpha , (j, m)} 
                                - \min_{m} \left[ A^{-1}_{\mathrm{L}} \right]_{\alpha , (j, m)} \right\}^{2}.               
               \end{eqnarray}
            \end{lemma}
            \textbf{Proof} (Lemma \ref{lemma:ErrorProb_LLS_LInfty})
               From the explicit form of the LLS estimate, $\bm{s}^{\mathrm{LLS}}_{N} = A^{-1}_{\mathrm{L}} ( \bm{f}_{N} - \bm{a}_{0})$, we have
               \begin{eqnarray}
                  \bm{s}^{\mathrm{LLS}}_{N} - \bm{s} = A^{-1}_{\mathrm{L}} \left\{ \bm{f}_{N} - \bm{p} (\rho) \right\}. \label{eq:sLLS-s}
               \end{eqnarray}               
               Let $m^{(j)}_{q}$ denote the $q$-th outcome in $n^{(j)}$ trials of the measurement $\bm{\Pi}^{(j)}$ ($j = 1, \ldots, J,\ q = 1, \ldots , n^{(j)}$).
               Then the $\alpha$-th element of the first term in the R. H. S. of equation (\ref{eq:sLLS-s}) is rewritten as
               \begin{eqnarray}
                  \left[ A^{-1}_{\mathrm{L}} \bm{f}_{N} \right]_{\alpha} 
                     &=& \sum_{j,m} \left[ A^{-1}_{\mathrm{L}} \right]_{\alpha, (j,m)} \frac{n^{(j)}_{m}}{n^{(j)}} \\     
                     &=& \sum_{j=1}^{J} \sum_{q=1}^{n^{(j)}}  \frac{\left[ A^{-1}_{\mathrm{L}} \right]_{\alpha, (j,m^{(j)}_{q})}}{n^{(j)}}. \label{eq:AinvL_sum}             
               \end{eqnarray}
               The R. H. S. of equation (\ref{eq:AinvL_sum}) is a sum of independent (and non-identical) random variables, and the expectation coincides with the second term in the R. H. S. of equation (\ref{eq:sLLS-s}).
               Each random variable in the sum takes value in 
               \begin{eqnarray}
                  \left[ \frac{1}{n^{(j)}}\min_{m} \left[ A^{-1}_{\mathrm{L}} \right]_{\alpha, (j,m)}, \frac{1}{n^{(j)}}\max_{m} \left[ A^{-1}_{\mathrm{L}} \right]_{\alpha, (j,m)} \right] .
               \end{eqnarray}             
               By applying Lemma \ref{lemma:HoeffdingTailIneq} to these random variables, we obtain
               \begin{eqnarray}
                 \Probability \left[ | s^{\mathrm{LLS}}_{N, \alpha } - s_{\alpha} |  > \delta \right]   
                 \le 2 \exp \left[ - 2 \delta^{2} / c^{\prime}_{\alpha}  \right], \label{eq:ErrorProb_LLS_alpha}      
               \end{eqnarray}
               where
               \begin{eqnarray}
                  c^{\prime}_{\alpha} 
                     &:=& \sum_{j, q} 
                               \left\{ \frac{ \max_{m} \left[ A^{-1}_{\mathrm{L}} \right]_{\alpha, (j,m)} }{n^{(j)}} 
                                       - \frac{ \min_{m} \left[ A^{-1}_{\mathrm{L}} \right]_{\alpha, (j,m)} }{n^{(j)}} \right\}^{2}  \notag  \\
                     &\phantom{:}=& \sum_{j} \frac{1}{n^{(j)}}
                                      \left\{ \max_{m} \left[ A^{-1}_{\mathrm{L}} \right]_{\alpha, (j,m)}  
                                       - \min_{m} \left[ A^{-1}_{\mathrm{L}} \right]_{\alpha, (j,m)}  \right\}^{2} \notag \\
                     &\phantom{:}=& \frac{c_{\alpha}}{N}. \label{eq:cAlphaPrime}                                                          
               \end{eqnarray}               
               By definition of the $\ell_{\infty}$-distance, $\| \bm{s}^{\mathrm{LLS}}_{N} - \bm{s} \|_{\infty} > \delta$ indicates that there exists at least one $\alpha \in \{ 1, \ldots , d^{2} -1 \}$ such that $| s^{\mathrm{LLS}}_{n, \alpha } - s_{\alpha} |  > \delta $.
               Then we have
               \begin{eqnarray}
                  \Probability \left[ \| \bm{s}^{\mathrm{LLS}}_{N} - \bm{s} \|_{\infty} > \delta \right]
                  &\le& \sum_{\alpha = 1}^{d^{2} -1}
                             \Probability \left[ | s^{\mathrm{LLS}}_{N, \alpha } - s_{\alpha} |  > \delta \right]. \label{eq:ErrorProb_LInfty_AVSum}                                                           
               \end{eqnarray}      
               By combining equations (\ref{eq:ErrorProb_LLS_alpha}), (\ref{eq:cAlphaPrime}), and (\ref{eq:ErrorProb_LInfty_AVSum}), we obtain equation (\ref{eq:ErrorProb_LLS_LInfty}).
            $\blacksquare$

            Finally, we prove Theorem 1 by combining Lemmas \ref{lemma:ErrorProb_ENM_LLS} and \ref{lemma:ErrorProb_LLS_LInfty}. \\       
            \textbf{Proof} (Theorem 1)
               Suppose that for two loss functions $\Delta$ and $\Delta^{\prime}$, there exists a constant $c>0$ such that $\Delta (\sigma , \sigma^{\prime} ) \le c \cdot  \Delta^{\prime} (\sigma , \sigma^{\prime } )$ holds for any Hermitian matrices $\sigma$ and $\sigma^{\prime}$. 
               Let $\rho^{\mathrm{est}}$ denote an estimator.
               Then for any $\rho \in \mathcal{S}(\mathcal{H})$,  $\Delta (\rho^{\mathrm{est}}_{N}, \rho) > \delta$ implies $\Delta^{\prime} (\rho^{\mathrm{est}}_{N}, \rho) > \delta / c$, and for arbitrary $\rho \in \mathcal{S} (\mathcal{H})$, estimator $\rho^{\mathrm{est}}$, set of positive integers $n^{(j)}$ satisfying $\sum_{j=1}^{J}n^{(j)} =N$, and positive number $\delta$,            
               \begin{eqnarray}
                  \Probability \left[ \Delta (\rho^{\mathrm{est}}_{N}, \rho) > \delta \right] 
                  \le  \Probability \left[ \Delta^{\prime} (\rho^{\mathrm{est}}_{N}, \rho) > \frac{\delta}{c}  \right] \label{eq:TwoDeltas_ErrorProb}             
               \end{eqnarray}   
               holds.
               The Hilbert-Schmidt distance and $\ell_{\infty}$-distance satisfy the following inequality \cite{GeometryofQuantumStates2006,MatrixAnalysisText1}:
               \begin{eqnarray}
                  \Delta^{\mathrm{HS}} ( \rho(\bm{s}^{\prime}) , \rho (\bm{s}) )
                  \le \frac{\sqrt{d^{2} - 1}}{2} \| \bm{s} - \bm{s}^{\prime} \|_{\infty}. \label{eq:HS_LInfty}
               \end{eqnarray}
               From equations (\ref{eq:TwoDeltas_ErrorProb}) and (\ref{eq:HS_LInfty}), we obtain               
               \begin{eqnarray}
                  \hspace{-5mm}\Probability \left[ \Delta^{\mathrm{HS}} (\rho^{\mathrm{LLS}}_{N}, \rho) > \delta \right] 
                  \le \Probability \left[ \| \bm{s}^{\mathrm{LLS}}_{N} - \bm{s} \|_{\infty} > \frac{2 \delta}{\sqrt{d^{2} - 1}} \right]. \label{eq:LLS_HS_LInfty} 
               \end{eqnarray}
               By combining equations (\ref{eq:Lemma2}), (\ref{eq:LLS_HS_LInfty}), and (\ref{eq:ErrorProb_LLS_LInfty}), we obtain
               \begin{eqnarray}
                  \hspace{-5mm}\Probability \left[ \Delta^{\mathrm{HS}} (\rho^{\mathrm{ENM}}_{N}, \rho) > \delta \right]           
                  \le 2 \sum_{\alpha = 1}^{d^{2} - 1} \exp \left[ -  \frac{8}{d^{2} - 1} \frac{1}{c_{\alpha}} \delta^{2} N \right].
               \end{eqnarray}
               The trace distance and the infidelity satisfy the following inequalities \cite{GeometryofQuantumStates2006,MatrixAnalysisText1}:
               \begin{eqnarray}
                  \Delta^{\mathrm{T}} (\rho^{\prime} , \rho ) &\le& \sqrt{\frac{d}{2}} \Delta^{\mathrm{HS}} (\rho^{\prime}, \rho ), \\
                  \Delta^{\mathrm{IF}} (\rho^{\prime}, \rho) &\le& \sqrt{2d} \Delta^{\mathrm{HS}} (\rho^{\prime}, \rho ), 
               \end{eqnarray}
               and from equation (\ref{eq:TwoDeltas_ErrorProb}) we obtain Theorem 1.   
            $\blacksquare$            
            
        Like quantum state tomography, quantum process tomography, which estimates quantum processes, is an important tool in quantum information experiments.
        A quantum process is described by a linear, trace-preserving, and completely positive map.
        By using the Choi-Jamiolkowski isomorphism \cite{Jamiolkowski72,Choi75}, this map can be represented by a density matrix.
        Therefore our result of quantum state tomography can be generalized to quantum process tomography straightforwardly.    

      In the proof of Lemma \ref{lemma:ErrorProb_ENM_LLS}, we proved that the estimation error probability of $\rho^{\mathrm{ENM}}$ is always smaller than that of $\rho^{\mathrm{LLS}}$ and the confidence level in Theorem 1 is about $\rho^{\mathrm{LLS}}$.
      Therefore a property of the projection in $\rho^{\mathrm{ENM}}$ is used in the proof, but the effect is not included in the confidence level in Theorem 1.
      The effect of the projection becomes larger as the prepared state is closer to the boundary of the state space.
      Usually it is expected that states close to the boundary would not effect on the value of the confidence level because a confidence level is derived by analyzing the state that is most difficult to estimate in all possible prepared states and it would be around the completely mixed state, which is far from the boundary.
      Therefore in many cases we could expect that the projection in $\rho^{\mathrm{ENM}}$ does not affect the confidence level.
      However, if we choose a loss function that put a high weight on states close to the boundary, like infidelity, the above expectation might not be true.
      The projection is a nonlinear map and the detailed analysis of the effect on the confidence level is a theoretically challenging problem.   

      In the proofs explained above, we used five inequalities for deriving the confidence level for the trace distance.
      The equalities of these inequalities do not hold simultaneously.
      This means that we under-evaluate the confidence levels of $\rho^{\mathrm{LLS}}$ and $\rho^{\mathrm{ENM}}$ and there must be possible improvement of Theorem 1.
      An improved confidence level guarantees a higher reliability of the estimation results for same data.
      This is an open problem that is important for theorists and experimentalists both.

      One interesting and natural question is how close the confidence level in Theorem 1 is from the optimal value.
      For this comparison, we need to calculate the optimal value of the confidence level.
      It seems to be very hard to analytically solve this problem, and a numerical analysis using a Monte Carlo method might be considered as a practical approach.
      In the analysis of the estimation error probability, however, we need to analyze rare events, $\{ \mbox{data}\ | \Delta (\rho^{\mathrm{ENM}}_{N}, \rho ) > \delta \}$, whose probability decreases exponentially fast with respect to $N$, and Monte Carlo methods require high computational cost in order to accurately evaluate the probability of such very rare events.
      Furthermore, in order to calculate the optimal confidence level, we need to perform the maximization of the estimation error probability over all possible states.  
      This optimization makes the computational cost higher.
      Therefore the comparison of our result to the optimal value is both a theoretically and also probably a numerically challenging open problem.
      On the other hand, the difficulty of the optimality analysis implies an advantage of our approach and result.
      That is, the confidence level in Theorem 1 might be far from the optimal value that is very hard to compute, but in exchange for optimality the confidence level in Theorem 1 is very easy to calculate.

\section{Confidence level for $k$-qubit state tomography}

   We derive the confidence level for a $k$-qubit state tomography explained in the manuscript.
   
   Suppose that we prepare $N$ identical copies of $\rho \in \mathcal{S} ( (\mathbbm{C}^{2})^{\otimes k})$ and make the three Pauli measurements with detection efficiency $\eta$ on each qubit.
   The POVMs describing the ideal Pauli measurements on each qubit, $\bm{\Pi}^{(i)} = \{ \Pi^{(i)}_{+1}, \Pi^{(i)}_{-1} \}$, are given as
   \begin{eqnarray}
      \Pi^{(i)}_{\pm 1} &:=& \frac{1}{2} \left( I \pm \bm{e}_{i} \cdot \bm{\sigma} \right),\\
   \end{eqnarray}
   where $i=1,2,3$, 
   \begin{eqnarray}
      \bm{e}_{1} := \left(
         \begin{array}{c}
         1 \\
         0 \\
         0 \\
         \end{array}
         \right), \
      \bm{e}_{2} := \left(
         \begin{array}{c}
         0 \\
         1 \\
         0 \\
         \end{array}
         \right),  \        
      \bm{e}_{3} := \left(
         \begin{array}{c}
         0 \\
         0 \\
         1 \\
         \end{array}
         \right), 
    \end{eqnarray}
    and
    \begin{eqnarray}
      \hspace{-5mm}
      \sigma_{1} := \left(
         \begin{array}{cc}
         0 & 1 \\
         1 & 0 
         \end{array}
         \right),  \     
      \sigma_{2} := \left(
         \begin{array}{cc}
         0 & -i \\
         i & 0 
         \end{array}
         \right),  \
      \sigma_{3} := \left(
         \begin{array}{cc}
         1 &  0\\
         0 & -1 
         \end{array}
         \right).                                   
   \end{eqnarray} 
   When the measurements have detention loss, the corresponding POVMs, $\bm{\Pi}^{\eta, (i)} = \{ \Pi^{\eta, (i)}_{+1}, \Pi^{\eta, (i)}_{-1}, \Pi^{\eta, (i)}_{0} \}$, are given as
   \begin{eqnarray}
      \Pi^{\eta, (i)}_{\pm 1} &:=& \frac{\eta}{2} \left( I \pm \bm{e}_{i} \cdot \bm{\sigma} \right)\\
      \Pi^{\eta, (i)}_{0} &:=& (1 - \eta ) I,
   \end{eqnarray}   
   where $\eta$ is the detection efficiency and takes the value from $0$ to $1$.
   The outcome ``$0$'' means no detection at the measurement trial.
   When we perform the imperfect Pauli measurements on each qubit, the POVM on $k$-qubit is given as
   \begin{eqnarray}
      \bm{\Pi}^{\eta, (\bm{i})} := \otimes_{q=1}^{k}\bm{\Pi}^{\eta, (i_{q})},
   \end{eqnarray} 
   where $\bm{i} = \{i_{q} \}_{q=1}^{k}$ and $i_{q} = 1, 2, 3$.
   The label of the different POVMs, $\bm{i}$, corresponds to $j$ in the manuscript. 
   Suppose that we perform each measurement described by $\bm{\Pi}^{\eta, (\bm{i})}$ equally $n:= N / 3^{k}$ times.
      
   Let us choose $\bm{\lambda}$ to be the set of tensor products of Pauli and identity matrices with the normalization factor $1 / \sqrt{2^{k-1}}$, {\it i.e.},
   \begin{eqnarray}
      \lambda_{\bm{\beta}} := \frac{1}{\sqrt{2^{k-1}}}\otimes_{q=1}^{k} \sigma_{\beta_{q}},
   \end{eqnarray}
   where $\bm{\beta} := \{ \beta_{q} \}_{q=1}^{k}$ and $\beta_{q} = 0, 1, 2, 3$.
   We eliminate from $\bm{\beta}$ the case that all $\beta_{q}$ are $0$.
   The label of the matrices, $\bm{\beta}$, corresponds to $\alpha$ in the manuscript.
   Using this $\bm{\lambda}$, any density matrices are represented as
   \begin{eqnarray}
      \rho = \frac{1}{2^{k}} I + \frac{1}{2} \bm{\lambda} \cdot \bm{s},
   \end{eqnarray}
   where 
   \begin{eqnarray}
      s_{\bm{\beta}} &=& \Trace \left[ \rho \lambda_{\bm{\beta}} \right] \\
                              &=& \frac{1}{\sqrt{2^{k-1}}}\Trace \left[ \rho \left( \otimes_{q=1}^{k} \sigma_{\beta_{q}} \right) \right]. \label{eq:Bloch_kqubit_2}
   \end{eqnarray}
   Equation (\ref{eq:Bloch_kqubit_2}) indicates that the parameter $s_{\bm{\beta}}$ is the expectation of a tensor product of ideal Pauli and identity matrices.
   
   In $k$-qubit state tomography with $k\ge 2$, we need to be careful about the treatment of multiple uses of same data. 
   For example, in order to estimate the expectation of $\sigma_{1} \otimes I$ in $2$-qubit case, we use the data of three types of measurements; $\sigma_{1} \otimes \sigma_{1}$, $\sigma_{1} \otimes \sigma_{2}$, and $\sigma_{1} \otimes \sigma_{3}$.
   Therefore the estimation of each parameter can de dependent even for $\rho^{\mathrm{LLS}}$.   
   
   We try to estimate these parameters from a data set of the imperfect Pauli measurements $\vec{\bm{\Pi}} := \{ \bm{\Pi}^{\eta, (\bm{i})} \}_{\bm{i}}$.   
   In order to calculate $c_{\alpha}$, we need to derive a matrix $B$ satisfying
   \begin{eqnarray}
      \bm{s} = B ( \bm{p} - \bm{a}_{0}).
   \end{eqnarray}
   This matrix $B$ corresponds to $A^{-1}_{\mathrm{L}}$ in the manuscript.
   Let $l$ denote the number of $I$ appearing in $\lambda_{\bm{\beta}}$.
   The number of $\lambda_{\bm{\beta}}$ including $l$ identities is $3^{k-l} \times \frac{k !}{l! (k-l)!}$.
   $\lambda_{\bm{\beta}} = I^{\otimes l} \otimes \left( \otimes_{q=l+1}^{k} \sigma_{i_{q}} \right) / \sqrt{2^{k-1}}$ is an example of such $\lambda_{\bm{\beta}}$.
   In this case, equation (\ref{eq:Bloch_kqubit_2}) is rewritten by the probability distributions of the imperfect Pauli measurement as 
   \begin{widetext}
   \begin{eqnarray}
      s_{\bm{\beta}} 
         &=& \frac{1}{\sqrt{2^{k-1}}} \sum_{m_{l+1}, \ldots , m_{k}; \atop m_{q} = \pm 1} \left( \prod_{q=l+1}^{k} m_{q} \right) 
                 p( m_{l+1}, \ldots , m_{k} | I^{\otimes l} \otimes (\otimes_{q=l+1}^{k} \bm{\Pi}^{(i_{q})}), \rho) \\
         &=& \frac{1}{\sqrt{2^{k-1}}} \sum_{m_{l+1}, \ldots , m_{k}; \atop m_{q} = \pm 1} \left( \prod_{q=l+1}^{k} m_{q} \right) 
          \frac{1}{\eta^{k-l}}p( m_{l+1}, \ldots , m_{k} | I^{\otimes l} \otimes (\otimes_{q=l+1}^{k} \bm{\Pi}^{\eta, (i_{q})}), \rho) \\                  
         &=& \frac{1}{\sqrt{2^{k-1}}} \sum_{i_{1}, \ldots , i_{l}; \atop i_{q} = 1, 2, 3} \sum_{m_{1}, \ldots , m_{l}; \atop m_{q} = \pm 1, 0} \sum_{m_{l+1}, \ldots , m_{k}; \atop m_{q} = \pm 1} \left( \prod_{q=l+1}^{k} m_{q} \right) \ \frac{1}{\eta^{k-l}} \frac{1}{3^{l}} p( m_{1}, \ldots , m_{k} | \bm{\Pi}^{\eta, (\bm{i})}, \rho). 
    \end{eqnarray}  
    \end{widetext}
    Therefore we have
    \begin{eqnarray}
       B_{\bm{\beta}, (\bm{i}, \bm{m})} = \pm \frac{1}{\sqrt{2^{k-1}}} \frac{1}{\eta^{k-l}} \frac{1}{3^{l}}, 
    \end{eqnarray}
    if $i_{q} = 1, 2, 3$ and $m_{q} = \pm 1, 0$ for $q= 1, \ldots , l$ and $i_{q} = \beta_{q}$ and $m_{q} = \pm 1$ for $q = l+1, \ldots , k$. 
    Otherwise $B_{\bm{\beta}, (\bm{i}, \bm{m})} = 0$.
    Then for each $\beta$ and $\bm{i}$, 
    \begin{eqnarray}
       &&\max_{\bm{m}} B_{\bm{\beta}, (\bm{i}, \bm{m})} -  \min_{\bm{m}} B_{\bm{\beta}, (\bm{i}, \bm{m})} \notag \\
       &=& \left\{
          \begin{array}{cl}
             \frac{2}{\sqrt{2^{k-1}} \eta^{k-l} 3^{l}} & \mbox{if}\ i_{q} = \beta_{q}\ \mbox{for}\ q = l+1, \ldots , k\\
             0 & \mbox{otherwise}            
          \end{array}
          \right. 
    \end{eqnarray}
    holds, and we obtain 
    \begin{eqnarray}
       c_{\bm{\beta}} 
       &=& \sum_{\bm{i}} 3^{k} \left\{ \max_{\bm{m}} B_{\bm{\beta}, (\bm{i}, \bm{m})} -  \min_{\bm{m}} B_{\bm{\beta}, (\bm{i}, \bm{m})} \right\}^{2} \\
       &=& \sum_{i_{1}, \ldots , i_{l}; \atop i_{q} = 1, 2, 3} 3^{k} \left\{ \frac{2}{\sqrt{2^{k-1}} \eta^{k-l} 3^{l}} \right\}^{2} \\
       &=& \frac{3^{k-l}}{2^{k-3} \eta^{2(k-l)}}.
    \end{eqnarray}
    From the above discussion, we can see that $c_{\bm{\beta}}$ takes same value for different $\lambda_{\bm{\beta}}$ with the same $l$. 
    The confidence level is calculated as
    \begin{eqnarray}
       \hspace{-10mm}CL (k) 
       &=& 1 - 2 \sum_{\bm{\beta}} \exp \left[ - \frac{b}{c_{\bm{\beta}}} \delta^{2} N \right] \notag \\
       &=& 1 - 2 \sum_{l=0}^{k-1} 3^{k-l} \binom{k}{l} \exp \left[ - b \frac{2^{k-3} \eta^{2(k-l)}}{3^{k-l}} \delta^{2} N \right].       
     \end{eqnarray} 
   When we choose the trace distance as the loss function, we have 
   \begin{eqnarray}
     b = \frac{16}{d (d^{2} -1 )} 
        = \frac{1}{2^{k-4} \cdot (2^{2k} - 1)}, 
   \end{eqnarray}
   and 
   \begin{eqnarray}
      \hspace{-12mm}CL(k) = 1 - 2 \sum_{l=0}^{k-1} 3^{k-l}  \binom{k}{l} \exp \left[ - \frac{2}{2^{2k} - 1} \frac{\eta^{2(k-l)}}{3^{k-l}} \delta^{2} N \right]. \label{eq:CL_kqubit_state_T}
   \end{eqnarray}   
   In one-qubit ($k=1$) and two-qubit ($k=2$) cases, we have
   \begin{eqnarray}
      CL (k=1) &=& 1 - 6 \exp \left[ - \frac{2}{9} \eta^{2} \delta^{2} N \right], \\                 
      CL (k=2) &=& 1 - 18 \exp \left[ - \frac{2}{135} \eta^{4} \delta^{2} N \right] \notag \\
      & & \hspace{2.5mm}- 12 \exp \left[ - \frac{2}{45} \eta^{2} \delta^{2} N \right].                        
   \end{eqnarray}
   
   As in the above discussion, when the directions of each Pauli measurement are perfectly orthogonal, it is easy to derive $c_{\bm{\beta}}$.
   When the directions are not orthogonal, we need to calculate $A^{-1}_{\mathrm{L}} = (A^{T}A)^{-1} A^{T}$.
   Then, it becomes more difficult to analyze $c_{\bm{\beta}}$, and we might need to calculate them numerically.

\section{Effect of systematic errors}

   Theorems 1 is valid for any informationally complete POVMs and is applicable for cases in which a systematic error exists.
   However, we must know exactly the mathematical representation of the systematic error in order to strictly verify a value of the confidence level.
   This assumption can be unrealistic in some experiments.
   In this section, we will weaken the assumption to a more realistic condition and give a formula of confidence levels in such a case.
   
   Let $\vec{\bm{\Pi}}$ denote a set of POVMs exactly describing the measurement used, and let $\vec{\bm{\Pi}}^{\prime} (\neq \vec{\bm{\Pi}})$ denote a set of POVMs that we mistake as the correct set of POVMs.
   We assume that $\vec{\bm{\Pi}}$ and $\vec{\bm{\Pi}}^{\prime}$ are both informationally complete.
   Suppose that we do not know $\vec{\bm{\Pi}}$, but we know that $\vec{\bm{\Pi}}$ is in a known set $\mathcal{M}$.
   For example, consider the case where an experimentalist wants to perform a projective measurement of $\sigma_{1}$.
   If they can guarantee that their actual measurement is prepared within $0.5$ degrees from the $x$-axis, and if their detection efficiency is $0.9$, then $\mathcal{M}$ is the set of all POVMs whose measurement direction and detection efficiency are within $0.5$ degrees of the $x$-axis and $0.9$, respectively. 

   For given relative frequencies $\bm{f}_{N}$, the correct and mistaken LLS estimates are 
   \begin{eqnarray}
      \bm{s}^{\mathrm{LLS}}_{N} &=& A^{-1}_{\mathrm{L}} (\vec{\bm{\Pi}}) \left\{ \bm{f}_{N} - \bm{a}_{0} \right\},\\
      \bm{s}^{\mathrm{LLS} \prime}_{N} &=& A^{-1}_{\mathrm{L}} (\vec{\bm{\Pi}}^{\prime}) \left\{ \bm{f}_{N} - \bm{a}_{0}^{\prime} \right\}.      
   \end{eqnarray}
   Then the actual and mistaken ENM estimates are 
   \begin{eqnarray}
      \bm{s}^{\mathrm{ENM}}_{N} &=& \argmin_{\bm{s}^{\prime} \in S} \| \bm{s}^{\prime} - \bm{s}^{\mathrm{LLS}}_{N} \|_{2}, \\
      \bm{s}^{\mathrm{ENM}\prime}_{N} &=& \argmin_{\bm{s}^{\prime} \in S} \| \bm{s}^{\prime} - \bm{s}^{\mathrm{LLS}\prime}_{N} \|_{2}.      
   \end{eqnarray} 
   Let $\rho^{\mathrm{ENM}}_{N}$ and $\rho^{\mathrm{ENM}\prime}_{N}$ denote the corresponding density matrix estimates.
   Let us define the size of the systematic error as
   \begin{eqnarray}
      \xi := \max_{\vec{\bm{\Pi}} \in \mathcal{M}} \Delta (\rho^{\mathrm{ENM}\prime}_{N}, \rho^{\mathrm{ENM}}_{N} ).      
   \end{eqnarray}
   This is a function of $\Delta$, $\bm{f}_{N}$, $\vec{\bm{\Pi}}^{\prime}$, and $\mathcal{M}$.
   Then for any $\rho \in \mathcal{S}(\mathcal{H})$ and $\vec{\bm{\Pi}} \in \mathcal{M}$,
   \begin{eqnarray}
      \Delta (\rho_{*}, \rho) 
         &\le& \Delta (\rho_{*}, \rho^{\mathrm{ENM}\prime}_{N}) 
              + \Delta (\rho^{\mathrm{ENM}\prime}_{N}, \rho^{\mathrm{ENM}}_{N} )   
              + \Delta (\rho^{\mathrm{ENM}}_{N}, \rho ) \notag \\
         &\le&  \Delta (\rho_{*}, \rho^{\mathrm{ENM}\prime}_{N}) 
              + \xi      
              + \delta \label{eq:ErrorThrethold_UnknownSystematicError}           
   \end{eqnarray}
   holds with probability at least
   \begin{eqnarray}
      \min_{\vec{\bm{\Pi}} \in \mathcal{M}} CL 
      = 1 - 2 \max_{\vec{\bm{\Pi}} \in \mathcal{M}} \sum_{\alpha = 1}^{d^{2} -1} \exp \left[ -\frac{b}{c_{\alpha}} \delta^{2} N \right]. \label{eq:ConfLevel_UnknownSystematicError}
   \end{eqnarray}   
   Using equations (\ref{eq:ErrorThrethold_UnknownSystematicError}) and (\ref{eq:ConfLevel_UnknownSystematicError}), we can evaluate the precision of state preparation, $\Delta (\rho_{*}, \rho )$, without knowing the true state $\rho$ and true sets of POVMs $\vec{\bm{\bm{\Pi}}}$.

\section{Effect of numerical errors}

   In this section, we analyze the effect of numerical errors and explain a method for evaluating the precision of the state preparation in the cases that numerical errors exits.
    
   The ENM estimator $\rho^{\mathrm{ENM}}$ requires a nonlinear minimization, which requires the use of  a numerical algorithm. 
   Suppose that we choose an algorithm for the minimization and obtain a result $\sigma^{\mathrm{ENM}}_{N}$ for a given data set. 
   In practice, there exists a numerical error on the result, and $\sigma^{\mathrm{ENM}}_{N}$ differs from the exact solution $\rho^{\mathrm{ENM}}_{N}$. 
   We cannot obtain the exact solution, but we can guarantee the accuracy of the numerical result with accuracy-guaranteed algorithms \cite{HighamText2002}.
   Suppose that we use an algorithm for which $\Delta (\sigma^{\mathrm{ENM}}_{N}, \rho^{\mathrm{ENM}}_{N}) \le \zeta$ is guaranteed. 
   Then
   \begin{eqnarray}
       \Delta (\rho_{*}, \rho ) 
       &\le& \Delta ( \rho_{*} , \sigma_{N}^{\mathrm{ENM}}) 
              + \Delta ( \sigma_{N}^{\mathrm{ENM}}, \rho^{\mathrm{ENM}}_{N}) 
              + \Delta ( \rho_{N}^{\mathrm{ENM}}, \rho)  \notag \\
      &\le& \Delta ( \rho_{*} , \sigma_{N}^{\mathrm{ENM}}) + \zeta + \delta \label{eq:ErrorThreshold_NumericalError}                
   \end{eqnarray}
   holds at the confidence level in Theorem 1.
   The error threshold is changed from $\delta$ to $\zeta + \delta$.

   Usually systematic and numerical errors both exists.
   In such a case, by combining equations (\ref{eq:ErrorThrethold_UnknownSystematicError}) and (\ref{eq:ErrorThreshold_NumericalError}), we can prove that the inequality
   \begin{eqnarray}
      \Delta (\rho_{*}, \rho)       
         \le \Delta (\rho_{*}, \sigma^{\mathrm{ENM}}_{N}) + \zeta + \xi + \delta
   \end{eqnarray}
   holds at the confidence level in equation (\ref{eq:ConfLevel_UnknownSystematicError}), where $\zeta$ is a numerical error threshold for $\vec{\bm{\Pi}}^{\prime}$.
   Therefore Theorem 1 with a modification can apply for the cases that systematic and numerical errors exist.

\section{Confidence level of constrained least squares estimator}

   In this section, we derive a confidence level for it using almost the same argument in Appendix A.
   By definition, the probability distribution of $\rho^{\mathrm{LLS}}_{N}$ is the projection of $\bm{f}_{N}$ on the probability space of trace-one Hermitian matrices ($\{ \bm{p}(\sigma ) | \sigma = \sigma^{\dagger}, \Trace [\sigma ] = 1 \}$), and we have
   \begin{eqnarray}
      \| \bm{p} (\rho^{\prime}) - \bm{f}_{N} \|_{2}^{\ 2} 
         &=& \| \bm{p} (\rho^{\prime}) - \bm{p} (\rho^{\mathrm{LLS}}_{N}) \|_{2}^{\ 2} \notag \\
         &&  + \| \bm{p} (\rho^{\mathrm{LLS}}_{N}) - \bm{f}_{N} \|_{2}^{\ 2}, \    \forall \rho^{\prime} \in \mathcal{S}(\mathcal{H}).
   \end{eqnarray}
   Therefore, Eq. (\ref{eq:CLS_def}) is rewritten as 
   \begin{eqnarray}
      \rho^{\mathrm{CLS}}_{N} 
      = \argmin_{\rho^{\prime} \in \mathcal{S}(\mathcal{H})} \| \bm{p} (\rho^{\prime}) - \bm{p} (\rho^{\mathrm{LLS}}_{N}) \|_{2}, \label{eq:CLS_LLS}
   \end{eqnarray}   
   and $\rho^{\mathrm{CLS}}_{N}$ is the projection of $\rho^{\mathrm{LLS}}_{N}$ on $\mathcal{S}(\mathcal{H})$ with respect to the $2$-norm on the probability space.
   We can see from Eqs. (\ref{def:ENM_state})  and (\ref{eq:CLS_LLS}) that $\rho^{\mathrm{ENM}}$ and $\rho^{\mathrm{CLS}}$ are the projections of $\rho^{\mathrm{LLS}}_{N}$ with respect to difference spaces (or different norms).
       
   Using Lemma \ref{lemma:Non-expandability}, we obtain
   \begin{eqnarray}
      \| \bm{p} (\rho^{\mathrm{CLS}}_{N}) - \bm{p} (\rho) \|_{2} 
      &\le& \| \bm{p} (\rho^{\mathrm{LLS}}_{N}) - \bm{p} (\rho) \|_{2}, \ \forall \rho \in \mathcal{S} (\mathcal{H}), \\      
      \| A ( \bm{s}^{\mathrm{CLS}}_{N} - \bm{s}) \|_{2} 
      &\le& \| A( \bm{s}^{\mathrm{LLS}}_{N} - \bm{s}) \|_{2}, \ \forall \bm{s} \in S,       
   \end{eqnarray}
   where $\bm{s}^{\mathrm{CLS}}_{N}$ is the Bloch vector corresponding to $\rho^{\mathrm{CLS}}_{N}$.
   Let us define $\| A \|_{\max}$ and $\| A \|_{\min}$ as 
   \begin{eqnarray}
      \| A \|_{\max} := \max_{\bm{v} \neq \bm{0}} \frac{\| A \bm{v} \|_{2}}{\| \bm{v} \|_{2}}, \\
      \| A \|_{\min} := \min_{\bm{v} \neq \bm{0}} \frac{\| A \bm{v} \|_{2}}{\| \bm{v} \|_{2}}.
   \end{eqnarray}
   When $\vec{\bm{\Pi}}$ is informationally complete, $A$ is full-rank and $\| A \|_{\min} > 0$.     
   We have
   \begin{eqnarray}
      \| A \|_{\min} \cdot \| \bm{s}^{\mathrm{CLS}}_{N} - \bm{s} \|_{2}
      &\le& \| A (\bm{s}^{\mathrm{CLS}}_{N} - \bm{s} ) \|_{2} \\
      &\le& \| A (\bm{s}^{\mathrm{LLS}}_{N} - \bm{s} ) \|_{2}  \\
      &\le& \| A \|_{\max} \cdot \| \bm{s}^{\mathrm{LLS}}_{N} - \bm{s} \|_{2}.        
   \end{eqnarray}
   We obtain
   \begin{eqnarray}
      \Probability \left[ \| \bm{s}^{\mathrm{CLS}}_{N} - \bm{s} \|_{2} > \delta \right] 
      \le \Probability \left[ \| \bm{s}^{\mathrm{LLS}}_{N} - \bm{s} \|_{2} > \frac{\| A \|_{\min}}{\| A \|_{\max}} \delta \right].      
   \end{eqnarray}
   From the same logic, we obtain the following theorem:
      \begin{theorem}[Confidence level of CLS estimator]\label{theorem:ConfLevel_CLS1}
         Under the same conditions and notation in Theorem \ref{theorem:ConfLevel_CLS_state},          
         \begin{eqnarray}
            \Delta (\rho^{\mathrm{CLS}}_{N} , \rho) \le \delta \label{eq:ErrorThrethold_CLS1}
         \end{eqnarray}
         holds with probability at least
         \begin{eqnarray}
            CL^{\mathrm{CLS}} :=1 - 2 \sum_{\alpha = 1}^{d^{2} - 1} \exp \left[- \left( \frac{\| A \|_{\min}}{\| A \|_{\max}} \right)^{2}\frac{b}{c_{\alpha}}\delta^{2} N \right], \label{eq:ConfidenceLevel_CLS1}
         \end{eqnarray}
      \end{theorem}       
      
   Compared to the confidence level of the ENM estimator (Eq. (\ref{eq:ConfidenceLevel_CLS})), there is an additional factor $\left( \frac{\| A \|_{\min}}{\| A \|_{\max}} \right)^{2} (\le 1)$ in the rate of exponential decrease in Eq. (\ref{eq:ConfidenceLevel_CLS1}).
   When $\| A \|_{\max} = \| A \|_{\min}$ holds, Eq. (\ref{eq:ConfidenceLevel_CLS1}) coincides with Eq. (\ref{eq:ConfidenceLevel_CLS}).
   Roughly speaking, the condition, $\| A \|_{\max} = \| A \|_{\min}$, implies that we perform measurements extracting information of each Bloch vector element with an equivalent weight.   
   When $\| A \|_{\max} > \| A \|_{\min}$, the confidence level of $\rho^{\mathrm{CLS}}$ is smaller than that of $\rho^{\mathrm{ENM}}$.
   This does not mean we can immediately conclude that $\rho^{\mathrm{CLS}}$ is less precise than $\rho^{\mathrm{ENM}}$ because their confidence levels,  Eqs. (\ref{eq:ConfidenceLevel_CLS}) and (\ref{eq:ConfidenceLevel_CLS1}), are probably not optimal as explained in the end of Appendix A.
   However, we can say that $\rho^{\mathrm{CLS}}$ is less precise than $\rho^{\mathrm{ENM}}$ insofar as Eqs. (\ref{eq:ConfidenceLevel_CLS}) and (\ref{eq:ConfidenceLevel_CLS1}) are the only confidence levels known for point estimators in quantum tomography to date.
   Additionally, the computational cost of $\rho^{\mathrm{ENM}}$ can be smaller than that of $\rho^{\mathrm{CLS}}$ as mentioned in the Letter.
   Therefore, we believe that the ENM estimator performs better than the CLS estimator and is at present our best choice.

\end{document}